\pdfoutput=1
\documentclass[aps,prapplied,reprint,superscriptaddress,amsmath,amssymb]{revtex4-2}

\usepackage{graphicx}
\usepackage{dcolumn}
\usepackage{bm}

\usepackage[utf8]{inputenc}
\usepackage[T1]{fontenc}
\usepackage{mathptmx}
\usepackage{etoolbox}

\usepackage{physics}

\usepackage[caption=false]{subfig}
\usepackage{soul}
\usepackage[normalem]{ulem}
\usepackage{xcolor}
\usepackage{tikz}
\usetikzlibrary{positioning, shapes, arrows}
\usepackage{multirow}
\usepackage{cancel}
\usepackage{float}

\begin{document}

\title{Fokker--Planck framework for stochastic octupole moment dynamics in chiral antiferromagnet Mn$_3$Sn}
\author{Siyuan Qian}
 \altaffiliation[Also at ]{Holonyak Micro and Nanotechnology Laboratory, University of Illinois, Urbana-Champaign, IL, USA.}

\author{Shaloo Rakheja}%
 \email{rakheja@illinois.edu}
\affiliation{ 
Holonyak Micro and Nanotechnology Laboratory, University of Illinois, Urbana-Champaign, IL, USA 
}%

\date{\today}

\begin{abstract}
We develop a reduced stochastic framework for thermally assisted octupole moment dynamics in Mn$_3$Sn by combining the reduced Landau--Lifshitz--Gilbert (LLG) equation with the Fokker--Planck formalism. The reduced model is benchmarked against the complete three-sublattice octupole dynamics and is shown to capture the essential switching behavior with good accuracy. We then derive the corresponding Fokker--Planck equation, which is implemented and solved via  
a CUDA-accelerated solver. 
The analysis shows that the octupole dynamics are highly sensitive to the out-of-plane grid resolution because ultrafast rotation of the octupole is controlled by its very small deviations from the basal plane. The solver is validated against Monte Carlo simulations through equilibrium distributions, relaxation trajectories, and switching times. Finally, we apply the method to thermally assisted field-driven switching and demonstrate efficient access to ultra-low error probabilities beyond the practical reach of direct Monte Carlo simulations.
\end{abstract}
\maketitle

\section{\label{sec:intro}Introduction}
\vspace{-10pt}
Mn$_3$Sn is now established as a prototypical noncollinear antiferromagnet in which magnetic symmetry, electronic topology, and collective order are strongly intertwined.~\cite{Nakatsuji2015,Kuroda2017,shukla2025spintronic} The material crystallizes in a hexagonal DO$_{19}$ structure and hosts an inverse triangular spin configuration on the kagome Mn sublattice.~\cite{Tomiyoshi1982,Park2018} Despite its antiferromagnetic nature, there is a 
residual net magnetic moment in Mn$_3$Sn
on the order of $10^{-3}~\mu_{\mathrm{B}}$ per Mn atom.~\cite{Nakatsuji2015} Mn$_3$Sn exhibits a large anomalous Hall response at room temperature, with anomalous Hall conductivity of the same order as that of ferromagnets, together with a sizable anomalous Nernst effect reaching approximately $0.35$--$0.6~\mu\mathrm{V\,K^{-1}}$ in the measured temperature range of $\sim 200$--$400~\mathrm{K}$.~\cite{Nakatsuji2015,Ikhlas2017} These transverse transport responses arise from substantial momentum-space Berry curvature generated by the noncollinear spin texture and its associated magnetic Weyl electronic structure.~\cite{Kuroda2017,Ikhlas2017} A unifying theoretical description of these properties is provided by the concept of ferroic cluster multipoles, in which the magnetic octupole serves as the primary order parameter governing anomalous transport and symmetry-breaking phenomena in Mn$_3$Sn.~\cite{Suzuki2017} This octupolar description has also proven useful for understanding domain behavior and its coupling to external perturbations,~\cite{Li2019} including direct imaging of octupole domains via the magneto-optical Kerr effect.~\cite{Higo2018}

The dynamics in Mn$_3$Sn can be described by the coherent rotation of the inverse triangular spin texture or, equivalently, the associated octupole order parameter. Small magnetic fields on the order of a few tenths of a Tesla
can deterministically reorient symmetry-related antiferromagnetic domains, leading to abrupt switching of the anomalous Hall response and demonstrating efficient coupling between the antiferromagnetic order and external fields despite the near zero net magnetization.~\cite{Nakatsuji2015,Sung2018APL} 
Mn$_3$Sn also exhibits current-driven dynamics~\cite{Deng2023,Xie2022,Nihei2025APL,shukla2024impact} mediated by spin-orbit torques (SOTs), 
enabling all-electrical switching at current densities of order 1--10\,MA/cm$^2$.~\cite{Tsai2020,Takeuchi2021} 
Experiments have revealed nanosecond electrical switching and terahertz-time-scale anomalous Hall dynamics in thin-film Mn$_3$Sn,~\cite{Matsuda2020,Deng2023,Xie2022} emphasizing the ultrafast intrinsic speed of this antiferromagnetic system. 
Recent temperature-dependent switching studies further show that the observed dynamics depend sensitively on thermal conditions, including the interplay among SOT, Joule heating, and thermal activation.~\cite{Nihei2025APL,Yoo2024} These quantitative features indicate that both deterministic torques and thermal fluctuations play essential roles in the dynamical response of Mn$_3$Sn.

Thermal behavior is central to the magnetic dynamics of $\mathrm{Mn_3Sn}$, yet both experimental~\cite{Sato2023,Kobayashi2023,Yoo2024} and theoretical~\cite{Konakanchi2025,Qian2025} studies remain in their infancy. Experimentally, Sato \emph{et al.}~\cite{Sato2023} investigated the Hall resistance of $\mathrm{Mn_3Sn}$ nanodots as a function of device diameter and magnetic-field strength, demonstrating that the retention characteristics can be engineered through geometric design. Although their work quantified thermal stability, 
the underlying description of the energy barrier and transition process was largely adapted from conventional ferromagnetic models, and thus the transition physics specific to noncollinear antiferromagnets remains insufficiently understood. 

In the context of current-driven switching, there is still an active debate over whether the dominant contribution arises from Joule heating or spin-orbit torque (SOT). Yoo \emph{et al.}~\cite{Yoo2024} reported, based on a model for Joule-heating-induced temperature rise, that the critical temperature for switching is relatively insensitive to external conditions and remains above the N\'eel temperature, while the threshold current density depends strongly on substrate properties and base temperature. By contrast, Nihei \emph{et al.}~\cite{Nihei2025APL} used temperature-dependent measurements over the range of 140--300~K and showed that the experimentally observed switching current is substantially lower than the current required to heat the film to the N\'eel temperature, supporting the conclusion that SOT is the dominant switching mechanism. Together, these contrasting findings point to an incomplete understanding of the thermal mechanisms governing magnetic switching in Mn$_3$Sn and emphasize the need for deeper theoretical exploration of thermally driven magnetic behavior.


Recent theoretical studies have mainly focused on the low-energy-barrier regime. Konakanchi \emph{et al.}~\cite{Konakanchi2025} theoretically investigated the thermal relaxation of the octupole order parameter in nanoscale chiral antiferromagnets. They identified distinct relaxation mechanisms in the regimes $\Delta E \ll k_\mathrm{B}T$ and $\Delta E > k_\mathrm{B}T$ where $\Delta E$ is the energy barrier without external field, and further analyzed the electrical tunability of the relaxation time. In our previous work,~\cite{Qian2025} we also presented a detailed analysis of the field-assisted escape and relaxation processes in $\mathrm{Mn_3Sn}$ for barriers in the range $2k_\mathrm{B}T \leq \Delta E \leq 6k_\mathrm{B}T$, and further proposed the potential applications of low-barrier Mn$_3$Sn in random number generation and probabilistic computing. To date, however, validation of these theoretical frameworks has relied primarily on Monte Carlo simulations,~\cite{GarciaPalacios1998,Evans2014} which require a large number of stochastic trajectories to achieve statistically reliable results. Moreover, existing theoretical studies remain largely limited to the low-barrier regime, motivating the development of an efficient numerical framework for investigating thermally driven octupolar dynamics. The Fokker--Planck equation provides a natural and well-established framework for this purpose, having been applied to thermal fluctuation problems in magnetic systems since the foundational work of Brown~\cite{Brown1963} and the comprehensive treatments of Risken~\cite{Risken1989} and Coffey \emph{et al.}~\cite{Coffey2012}

In this work, we develop a stochastic framework for studying the octupolar dynamics of Mn$_3$Sn. Starting from the reduced stochastic Landau--Lifshitz--Gilbert (LLG) equation for the octupole moment, which we validate against the complete three-sublattice stochastic LLG model for Mn$_3$Sn, we derive the corresponding Fokker--Planck equation and implement a CUDA-accelerated numerical solver. Using this framework, we identify the narrow out-of-plane region that governs the ultrafast dynamics of the octupole moment, establish the grid-resolution requirements needed for accurate simulation, and benchmark the Fokker--Planck results against Monte Carlo simulations. We then apply the method to thermally assisted, field-driven switching and show that it enables efficient evaluation of equilibrium distributions, switching times, and ultra-low error probabilities. Finally, we compare the computational efficiency of the Fokker--Planck and Monte Carlo approaches, demonstrating the clear advantage of the Fokker--Planck method for efficient simulation of stochastic octupolar dynamics.

\vspace{-10pt}
\section{\label{sec:equation}Landau--Lifshitz--Gilbert equation analysis}
\begin{figure}
    \centering
    \includegraphics[width=0.95\linewidth]{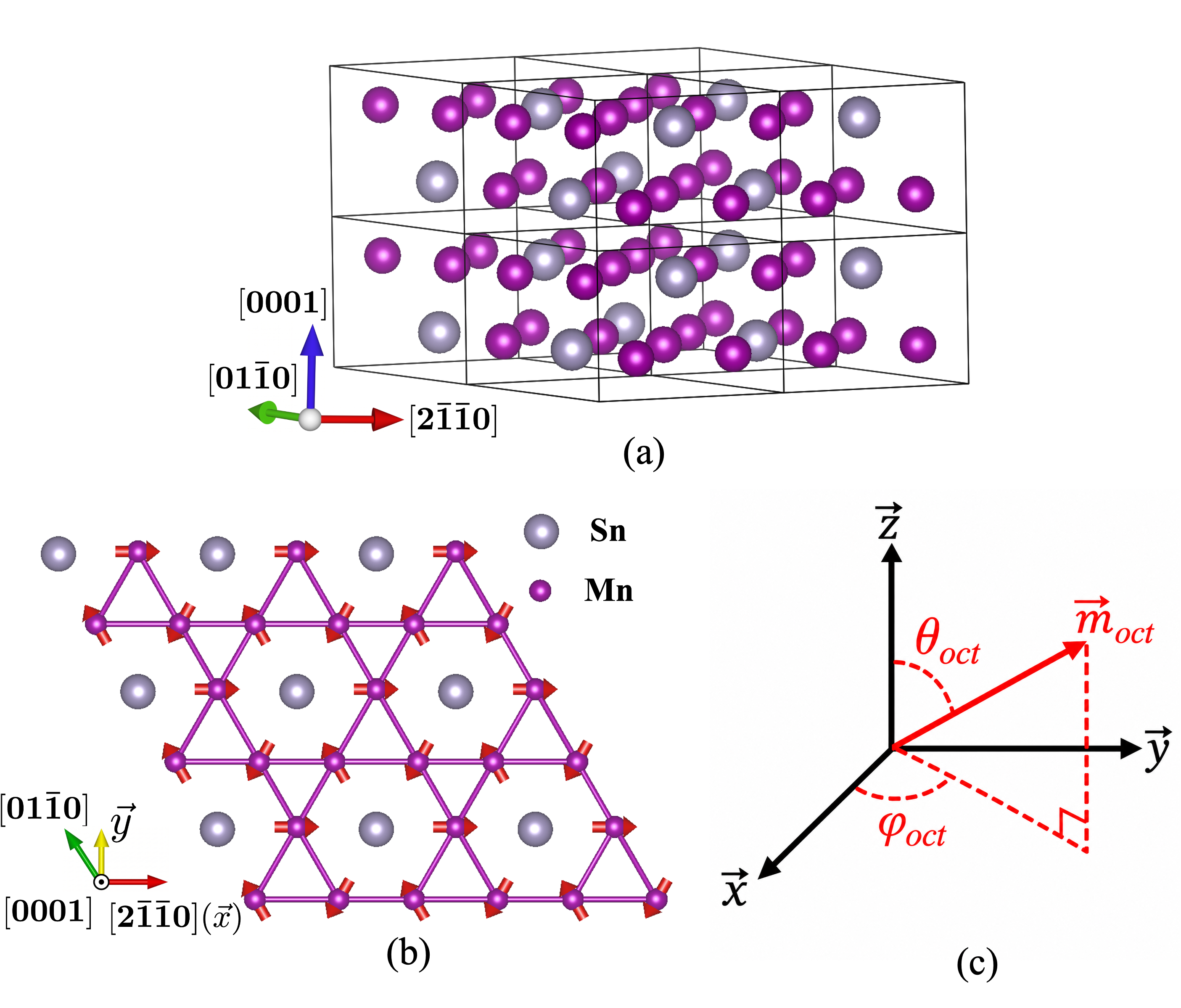}
    \caption{(a) Crystal structure of $\mathrm{Mn_3Sn}$. (b) Kagome plane of $\mathrm{Mn_3Sn}$ and sublattice magnetic order. (c) The polar($\theta_\mathrm{oct}$) and azimuthal($\varphi_\mathrm{oct}$) angles of $\boldsymbol{m}_\mathrm{oct}$.}
    \label{fig:structure}
\end{figure}

\vspace{-10pt}
Mn$_3$Sn crystallizes in the hexagonal DO$_{19}$ structure (space group $P6_3/mmc$) as depicted in Fig.~\ref{fig:structure}(a).~\cite{osti_1197606,momma2011vesta} Its primitive unit cell contains six Mn atoms forming two stacked Kagome layers, separated by Sn planes along the $c$ axis. Within each kagome layer, the Mn moments occupy three symmetry-equivalent sublattice sites, defining a natural three-sublattice basis for the magnetic degrees of freedom. Strong antiferromagnetic exchange stabilizes a noncollinear inverse-triangular spin configuration in which the three sublattice moments lie predominantly in the basal plane and are oriented $120^\circ$ apart as shown in Fig.~\ref{fig:structure}(b). Spin--orbit coupling, magnetocrystalline anisotropy, and Dzyaloshinskii--Moriya interaction (DMI) select the in-plane orientation and chirality of this texture. Although time-reversal symmetry is broken, the vector sum of the three sublattice moments nearly vanishes, motivating a description in terms of a collective octupole order parameter.

To represent the magnetic state compactly, we introduce the cluster magnetic octupole moment,
\begin{equation}
    \boldsymbol{m}_\mathrm{oct}  = \frac{1}{3} \mathcal{M}_{zx} \left[ R\left(\frac{2\pi}{3}\right) \boldsymbol{m}_1 + R\left(-\frac{2\pi}{3}\right) \boldsymbol{m}_2 + \boldsymbol{m}_3 \right],
\end{equation}
where $\mathcal{M}_{zx}$ denotes reflection about the $zx$ plane and $R$ denotes an anticlockwise rotation about the $z$ axis and $\boldsymbol{m}_i$ is the i-th sublattice in a unit cell. The octupole moment $\boldsymbol{m}_\mathrm{oct}$ is commonly described in spherical coordinates by its polar angle $\theta_\mathrm{oct}$ and azimuthal angle $\varphi_\mathrm{oct}$, as illustrated in Fig.~\ref{fig:structure}(c).

\begin{figure*}
    \centering
    \includegraphics[width=0.95\linewidth]{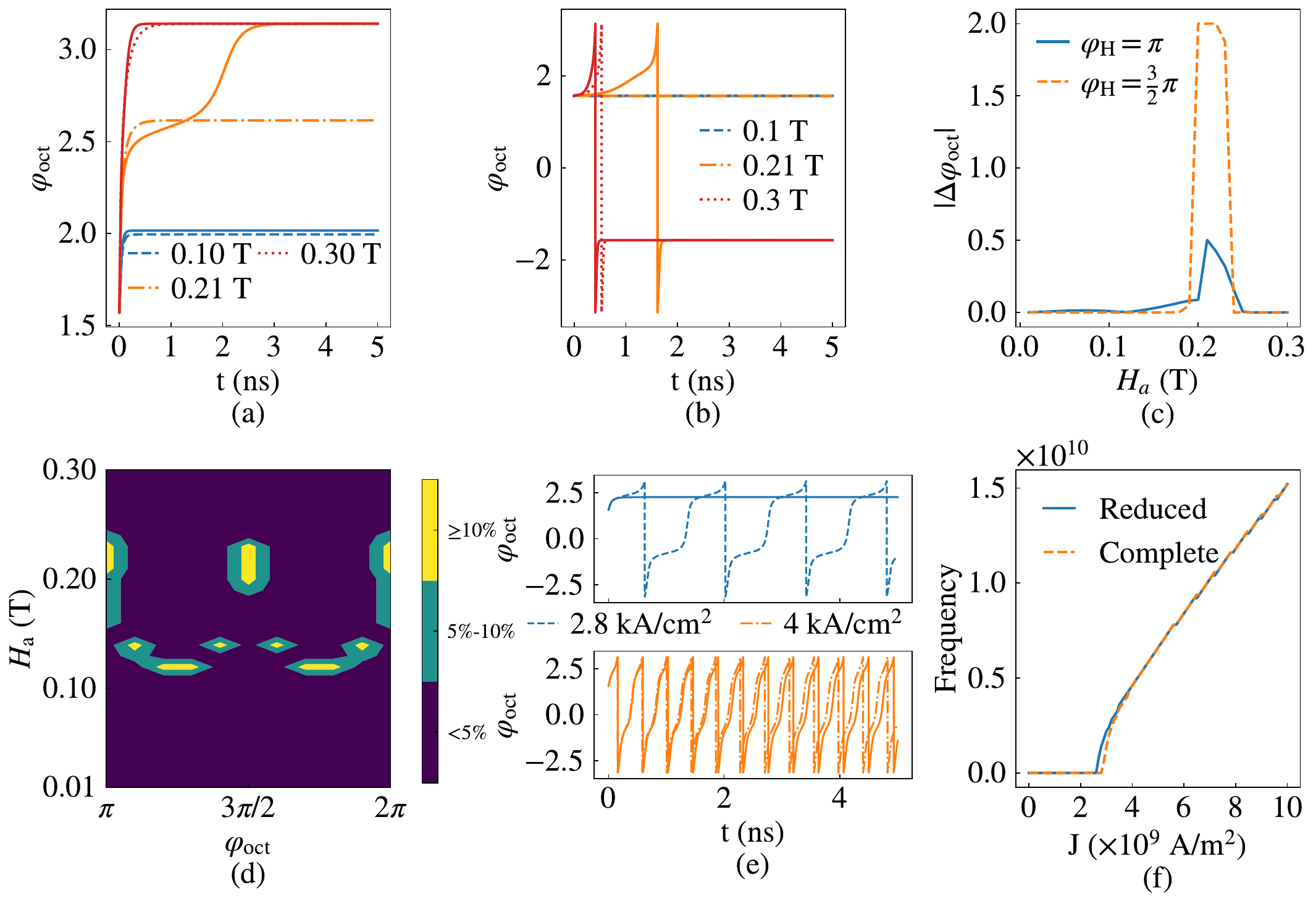}
    \caption{Comparison of the reduced model with the complete LLG simulation. Solid lines denote the complete LLG results. (a,b) Time evolution of the octupolar angle $\varphi_{\mathrm{oct}}$ under applied fields of $0.10$, $0.21$, and $0.30$ T for (a) $\varphi_{\mathrm{H}}=\pi$ and (b) $\varphi_{\mathrm{H}}=3\pi/2$. (c) Absolute error of the reduced model relative to the complete LLG simulation as $H_{\mathrm{a}}$ is swept from $0$ to $0.30$ T. (d) Relative error of the reduced model in the $(H_{\mathrm{a}},\varphi_{\mathrm{H}})$ parameter space for $0.01 \le H_{\mathrm{a}} \le 0.30$ T and $\pi \le \varphi_{\mathrm{H}} \le 2\pi$. (e) Time evolution of $\varphi_{\mathrm{oct}}$ for two spin-current magnitudes. (f) Oscillation frequency versus current magnitude for the reduced and complete LLG models.}
    \label{fig:reduced comparison}
\end{figure*}

An equation of motion that describes the spin dynamics of Mn$_3$Sn purely in terms of $\boldsymbol{m}_\mathrm{oct}$ can be derived from the Lagrange equation,
\begin{equation}
\frac{\partial \mathcal{L}}{\partial q_i}-\frac{\mathrm{d}}{\mathrm{d}t}\frac{\partial \mathcal{L}}{\partial \dot{q}_i}=\frac{\partial \mathcal{W}}{\partial \dot{q}_i}\qquad (q_i=\theta_\mathrm{oct},\,\varphi_\mathrm{oct}).
\end{equation}
The Rayleigh dissipation functional, $\mathcal{W}$, includes the Gilbert damping contribution $\mathcal{W}_\mathrm{damp} = \frac{\alpha M_s }{2\gamma} \sum_{i=1}^3 \left(\dot{\theta}_i^2 + \sin^2\theta_i\,\dot{\varphi}_i^2\right)$ and the spin-driven contribution $\mathcal{W}_\mathrm{spin} = \frac{M_s \tau}{\gamma}\,\boldsymbol{\sigma}\cdot(\dot{\boldsymbol{m}}_i \times \boldsymbol{m}_i)$. The Lagrangian is $\mathcal{L} = \mathcal{L}_\mathrm{B} - \mathcal{H}$, with Berry-phase term $\mathcal{L}_B = \sum_{i = 1,2,3}\frac{M_s}{\gamma} (1-\cos \theta_i)\dot{\varphi}_i$. The Hamiltonian of the three-sublattice unit-cell model, including a thermal field and written in terms of the sublattice moments, is
\begin{equation}
    \begin{aligned}
        \mathcal{H}(\boldsymbol{m}) &= J_\mathrm{E}((1+\delta_\mathrm{E})\boldsymbol{m}_1 \cdot \boldsymbol{m}_2+\boldsymbol{m}_2\cdot \boldsymbol{m}_3+\boldsymbol{m}_3\cdot \boldsymbol{m}_1)\\
        &+D_\mathrm{M} \boldsymbol{z}\cdot (\boldsymbol{m}_1\times \boldsymbol{m}_2+\boldsymbol{m}_2\times \boldsymbol{m}_3+\boldsymbol{m}_3\times \boldsymbol{m}_1)\\
        &-\sum_{i=1}^{3}{\left[K_\mathrm{u}(\boldsymbol{m}_i\cdot \boldsymbol{u}_i)^2+M_\mathrm{s} \boldsymbol{H}_\mathrm{a}\cdot \boldsymbol{m}_i)\right]}\\
        &-\sum_{i=1}^{3}{M_\mathrm{s} \boldsymbol{m}_i \cdot \boldsymbol{H}_{\mathrm{th},i}},
    \end{aligned}
    \label{eq:energy complete}
\end{equation}
where $\boldsymbol{m}_1$, $\boldsymbol{m}_2$, and $\boldsymbol{m}_3$ denote the magnetic moments of the three sublattices. The parameters $\delta_\mathrm{E}$, $M_{\mathrm{s}}$, $J_{\mathrm{E}}$, $D_{\mathrm{M}}$, $K_{\mathrm{u}}$, and $\boldsymbol{u}_i$ are, respectively, the dimensionless strain strength, the saturation magnetization of each sublattice, the symmetric exchange constant, the DMI strength, the single-ion uniaxial magnetocrystalline anisotropy constant, and the uniaxial easy axis of the $i$-th sublattice. The thermal field on the $i$-th sublattice, $\boldsymbol{H}_{\mathrm{th},i}$, has mean $\langle\boldsymbol{H}_{\mathrm{th},i}(t)\rangle = 0$ and correlation $\langle\boldsymbol{H}_{\mathrm{th},i}(t)\boldsymbol{H}_{\mathrm{th},j}(t')\rangle = \frac{2\alpha k_\mathrm{B} T}{\gamma \mu_0 M_\mathrm{s}^2 \mathcal{V}}\delta_{ij}\delta(t-t')$.

Using perturbation theory,~\cite{Qian2025,Konakanchi2025,shukla2024impact} the Lagrangian can be recast solely in terms of $\boldsymbol{m}_\mathrm{oct}$ as
\begin{subequations}
    \begin{equation}
        \begin{aligned}
            \mathcal{L}_\mathrm{B} = -\frac{3M_s}{\gamma} (1-\cos \theta_\mathrm{oct})\dot{\varphi}_\mathrm{oct},
        \end{aligned}
    \label{LB moct}
    \end{equation}
    \begin{equation}
        \begin{aligned}
            &\mathcal{H}(\theta_\mathrm{oct}, \varphi_\mathrm{oct}) = \frac{3}{4}M_\mathrm{s} H_\mathrm{K} \cos(2\varphi_\mathrm{oct}) + \frac{3}{2}M_\mathrm{s} H_\mathrm{J} \cos^2 \theta_\mathrm{oct}\\
            &-\frac{M_\mathrm{s} H_\mathrm{a} \sin(\theta_\mathrm{H})}{J_\mathrm{E}+\sqrt{3}D_\mathrm{M}}\Big[ K_\mathrm{u} \cos(\varphi_{\mathrm{oct}}-\varphi_{\mathrm{H}})+J_\mathrm{E}\delta_E\cos(\varphi_{\mathrm{oct}}+\varphi_{\mathrm{H}})\Big] \\
            &-3 M_\mathrm{s} \boldsymbol{m}_\mathrm{oct} \cdot \boldsymbol{H}_{th,\mathrm{oct}},
        \end{aligned}
    \label{Hamiltonian moct}
    \end{equation}
    \begin{equation}
        \begin{aligned}
            \mathcal{W} &= \frac{3\alpha M_s }{2\gamma}\left(\dot{\theta}_\mathrm{oct}^2 + \sin^2\theta_\mathrm{oct}\,\dot{\varphi}_\mathrm{oct}^2\right)\\
            &+\frac{3M_s }{\gamma}\,\tau\left(\dot{\theta}_\mathrm{oct}\,\Gamma_{\varphi}(-\varphi_\mathrm{oct})+\sin\theta_\mathrm{oct}\,\dot{\varphi}_\mathrm{oct}\,\Gamma_{\theta}(-\varphi_\mathrm{oct})\right),
        \end{aligned}
    \label{dissipation moct}
    \end{equation}
\end{subequations}
where $H_\mathrm{J} = \frac{(3J_\mathrm{E}+\sqrt{3}D_\mathrm{M})}{M_s}$ and $H_\mathrm{K} = -\frac{(4K_\mathrm{u} J_\mathrm{E}\delta_\mathrm{E})}{3M_s(J_\mathrm{E}+\sqrt{3}D_\mathrm{M})}$ are the exchange-field strength and uniaxial anisotropy-field strength, respectively. The SOT amplitude is $\tau = \frac{\hbar J}{2 \mu_0 e M_\mathrm{s}d_z}$, while $\Gamma_\varphi$ and $\Gamma_\theta$ are the $\varphi$ and $\theta$ components of the torque, respectively and $d_z$ is the thickness of $\mathrm{Mn_3Sn}$. Here, $\theta_\mathrm{H}$ and $\varphi_\mathrm{H}$ denote the polar and azimuthal angles of $\boldsymbol{H}_\mathrm{a}$, respectively, while $\theta_\mathrm{H}$ and $\varphi_\mathrm{H}$ specify the polar and azimuthal angles of the applied field $\boldsymbol{H}_\mathrm{a}$. It is worth noting that the thermal field $\boldsymbol{H}_\mathrm{th,oct}$ in Eq.~(\ref{Hamiltonian moct}) satisfies the correlation
$\langle \boldsymbol{H}_\mathrm{th,oct,i}(t)\boldsymbol{H}_\mathrm{th,oct,j}(t')\rangle = \frac{2\alpha k_\mathrm{B} T}{3\gamma \mu_0 M_\mathrm{s}^2 \mathcal{V}}\delta_{ij}\delta(t-t')$,
so its root-mean-square magnitude is $1/\sqrt{3}$ of that acting on each sublattice.

Consequently, a closed set of equations governing the dynamics of $\boldsymbol{m}_\mathrm{oct}$ is obtained as
\begin{subequations}
    \begin{equation}
        \begin{aligned}
            (1+\alpha^2)\,\dot{\theta}_\mathrm{oct} &= \gamma\left(\alpha H_{\theta_\mathrm{oct}}-H_{\varphi_\mathrm{oct}}\right) \\
            &+ \tau\left(\Gamma_{\theta}(\varphi_\mathrm{oct}) - \alpha\,\Gamma_{\varphi}(-\varphi_\mathrm{oct})\right),
        \end{aligned}
    \end{equation}
    \begin{equation}
        \begin{aligned}
            (1+\alpha^2)\,\sin \theta_\mathrm{oct}\,\dot{\varphi}_\mathrm{oct} &= \gamma\left(H_{\theta_\mathrm{oct}}+ \alpha H_{\varphi_\mathrm{oct}}\right) \\
            &- \tau\left(\alpha\,\Gamma_{\theta}(\varphi_\mathrm{oct}) + \Gamma_{\varphi}(-\varphi_\mathrm{oct})\right).
        \end{aligned}
    \end{equation}
    \label{eq: reduced}
\end{subequations}
The equations are referred to as Reduced LLG equations for the remainder of this paper and additional derivation details are provided in the Supplementary Material.

We numerically solve Eq.~(\ref{eq: reduced}) and benchmark the results against direct numerical integration of the coupled stochastic LLG equations~\cite{Qian2025}. The full dynamics are governed by
$
\frac{\partial \boldsymbol{m}_i}{\partial t}
=
-\gamma \left(\boldsymbol{m}_i \times \boldsymbol{H}_i^{\mathrm{eff}}\right)
+
\alpha \left(\boldsymbol{m}_i \times \frac{\partial \boldsymbol{m}_i}{\partial t}\right),
$
where the effective field is given by
$
\boldsymbol{H}_i^{\mathrm{eff}}
=
-\frac{1}{M_\mathrm{s}}
\frac{\partial \mathcal{H}}{\partial \boldsymbol{m}_i},
$
and is obtained from the full Hamiltonian in Eq.~(\ref{eq:energy complete}). Figures~\ref{fig:reduced comparison}(a) and (b) compare the time evolution of $\varphi_\mathrm{oct}$ predicted by the reduced model and the complete LLG equations for field directions $\varphi_\mathrm{H}=\pi$ and $\varphi_\mathrm{H}=3\pi/2$, respectively. Overall, the reduced equations reproduce the full LLG trajectories very well over nearly the entire field range, with noticeable deviations only within a narrow window around $H_\mathrm{a}\approx 0.21~\mathrm{T}$. As shown more clearly in Fig.~\ref{fig:reduced comparison}(c), this discrepancy corresponds to a pronounced displacement of the equilibrium state in that region.

It should be noted that the maximum final-state difference $\Delta \varphi_\mathrm{oct}$ for different field directions is not directly comparable as an error metric. For $\varphi_\mathrm{H}=\pi$, the equilibrium state shifts gradually from $\pi/2$ toward $\pi$ as the field increases, whereas for $\varphi_\mathrm{H}=3\pi/2$ the system undergoes a $180^\circ$ switching event once the threshold field is exceeded. To account for this difference in total angular excursion, we define the relative error as
$
\left | \frac{\varphi_\mathrm{com,final}-\varphi_\mathrm{red,final}}
{\varphi_\mathrm{com,final}-\varphi_\mathrm{com,ini}} \right |,
$
where ``com'' and ``red'' denote the complete and reduced models, respectively, and ``ini'' and ``final'' denote the initial and final states. We then sweep $\varphi_\mathrm{H}$ from $\pi$ to $2\pi$ and $H_\mathrm{a}$ from $0.01~\mathrm{T}$ to $0.3~\mathrm{T}$. The resulting error map in Fig.~\ref{fig:reduced comparison}(d) shows that most parameter combinations yield a relative error below $5\%$, confirming that the reduced model remains accurate across a broad field range.

A similarly small mismatch is observed in the current-driven oscillation regime, as shown in Figs.~\ref{fig:reduced comparison}(e) and (f). Near the threshold current density, for example at $J=2.8~\mathrm{kA/cm^2}$, the reduced LLG model does not sustain oscillation as readily as the complete LLG model. Nevertheless, Fig.~\ref{fig:reduced comparison}(f) shows that the predicted oscillation frequency agrees closely with the complete-model result over almost the entire current range, and the disagreement is confined to a very narrow region near threshold.

At present, the origin of the mismatch between the reduced and complete LLG equations has not been fully analyzed. Although the final-state difference in $\varphi_\mathrm{oct}$ can be as large as about $2.0$ in the mismatch region, this occurs where the energy barrier is approximately zero. In such a regime, the discrepancy likely reflects information lost during the perturbative reduction of the full Hamiltonian. As shown in the Supplementary Material, the Hamiltonian is expanded only to second order, and several higher-order contributions are neglected to obtain a tractable analytical expression for the equilibrium state. Despite these small discrepancies, the reduced LLG equation remains effective for describing the dynamics of $\boldsymbol{m}_\mathrm{oct}$. Further validation in the presence of thermal noise will be carried out using the Fokker--Planck formalism.

\begin{figure*}
    \centering
    \includegraphics[width=0.95\linewidth]{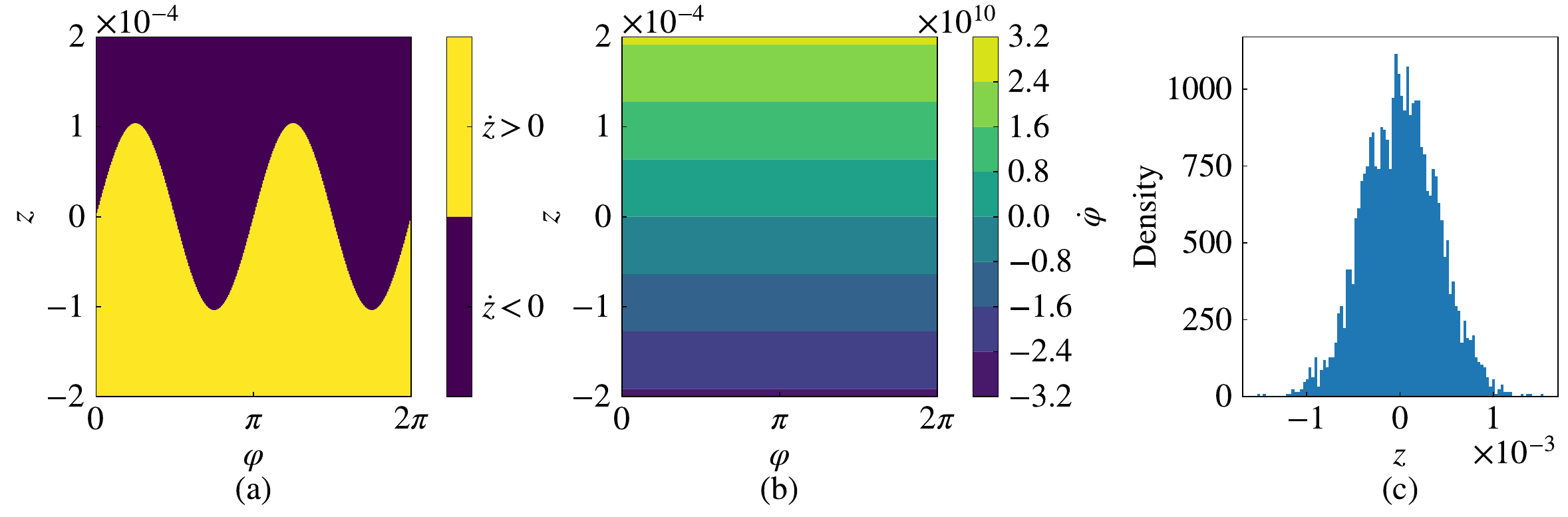}
    \caption{(a) Binary map of $\dot{z}$ in the range $-2\times10^{-4}\le z\le 2\times10^{-4}$ and $0\le \varphi_\mathrm{oct}\le 2\pi$, where purple denotes negative values and yellow denotes positive values. (b) Distribution of $\dot{\varphi}_\mathrm{oct}$ over the same range, $-2\times10^{-4}\le z\le 2\times10^{-4}$ and $0\le \varphi_\mathrm{oct}\le 2\pi$. (c) Distribution of $m_{\mathrm{oct},z}$ obtained from thermal-assisted reduced LLG simulations at $T=300~\mathrm{K}$ with an energy barrier of $\Delta E=2~k_\mathrm{B}T$.}
    \label{fig:zdot}
\end{figure*}
\vspace{-10pt}
\section{Fokker--Planck analysis}
\vspace{-10pt}
Instead of performing Monte Carlo sampling of the stochastic LLG equation, the Fokker--Planck equation provides an equivalent probabilistic description of the spin dynamics by governing the time evolution of the probability density over the angular phase space. Within this framework, deterministic effective fields and torques give rise to drift terms, while thermal fluctuations enter through diffusion terms whose strength is determined by the fluctuation--dissipation relation. This formulation is particularly advantageous for systematically analyzing relaxation dynamics, switching statistics, and nonequilibrium steady states under damping and field- or current-driven spin torques.

The Fokker--Planck equation for the probability density function $P(\theta,\varphi)$ is written as
\begin{subequations}
\begin{equation}
\frac{\partial P}{\partial t}
= -\nabla \cdot \mathcal{J}
= -\frac{1}{\sin\theta}\,\partial_\theta\!\left(\sin\theta\, \mathcal{J}_\theta\right)
-\frac{1}{\sin\theta}\,\partial_\varphi \mathcal{J}_\varphi,
\end{equation}
\begin{equation}
\mathcal{J}_\theta = \dot{\theta}P - D\frac{\partial P}{\partial \theta},
\qquad
\mathcal{J}_\varphi = \dot{\varphi}P - \frac{D}{\sin\theta}\frac{\partial P}{\partial \varphi},
\end{equation}
\end{subequations}
where $\mathcal{J}$ denotes the probability flux. The diffusion constant is given by
\begin{equation}
D = \frac{\alpha \gamma k_B T}{3(1+\alpha^2)\,M_s \mathcal{V}},
\end{equation}
which is reduced by a factor of $1/3$ relative to the ferromagnetic case, consistent with the reduced thermal-field correlation discussed previously.

To solve the equation numerically, we introduce the variable $z=\cos\theta$ and decompose the time evolution into drift and diffusion operators,
\begin{equation}
\frac{\partial P}{\partial t} = \mathcal{A}(P) + \mathcal{D}(P),
\end{equation}
where
\begin{equation}
\mathcal{A}(P)
= -\partial_z(\dot{z}P)
-\frac{1}{\sqrt{1-z^2}}\,\partial_\varphi(\dot{\varphi}P)
\end{equation}
represents the deterministic drift, and
\begin{equation}
\mathcal{D}(P)
= \partial_z\!\left(D(1-z^2)\partial_z P\right)
+ \partial_\varphi\!\left(\frac{D}{1-z^2}\partial_\varphi P\right)
\end{equation}
accounts for thermal diffusion. Using a first-order operator-splitting scheme, we treat the drift term explicitly and the diffusion term implicitly:
\begin{equation}
\left(I-\Delta t\,\mathcal{D}\right)P^{n+1}
= P^n+\Delta t\,\mathcal{A}(P^n)
\equiv P^{*}.
\end{equation}

We then apply a Fourier transform along the $\varphi$ direction, under which $\partial^2 P/\partial \varphi^2$ maps to $-m^2 P_m$, where $P_m$ denotes the $m$th Fourier mode of $P$. The implicit diffusion step is thereby reduced to a tridiagonal system along the $z$ direction for each Fourier mode $m$:
\begin{equation}
\begin{aligned}
&P^{n+1}_{i,j,m}
-\frac{D\Delta t}{\Delta z^2}
\Big[
(1-z_{i+1/2}^2)\left(P^{n+1}_{i+1,j,m}-P^{n+1}_{i,j,m}\right) \\
&
-(1-z_{i-1/2}^2)\left(P^{n+1}_{i,j,m}-P^{n+1}_{i-1,j,m}\right)
\Big]+ \frac{D\Delta t\, m^2}{1-z_i^2}\,P^{n+1}_{i,j,m}
= P^{*}_{i,j,m}.
\end{aligned}
\end{equation}
Based on this formulation, we develop a CUDA implementation of the Fokker--Planck solver so that the computation can be efficiently accelerated through parallel processing. For the benchmark, we also conduct Monte Carlo simulations based on Reduced LLG equations and Complete LLG equations with ensemble size 4096.


\begin{figure*}
    \centering
    \includegraphics[width=0.95\linewidth]{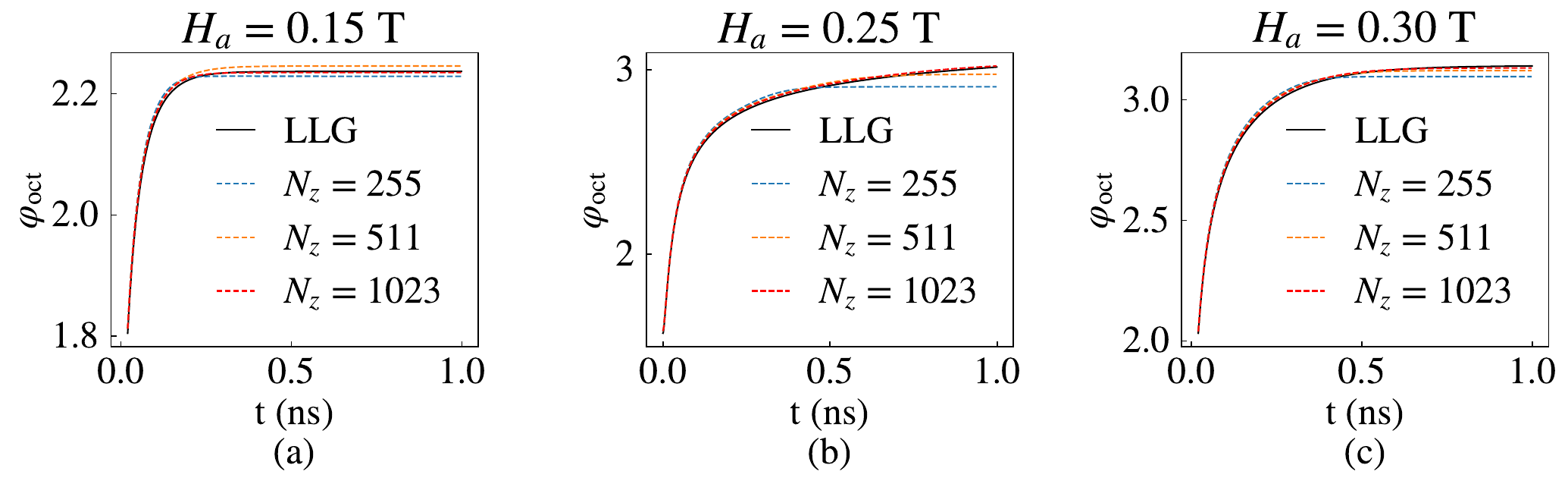}
    \caption{Deterministic benchmark of the Fokker--Planck solver against the reduced LLG equation for different grid resolutions in the $z$ direction. The time evolution of $\varphi_\mathrm{oct}$ is shown for (a) $H_a=0.15~\mathrm{T}$, (b) $H_a=0.25~\mathrm{T}$, and (c) $H_a=0.30~\mathrm{T}$. The black solid curves denote the reduced LLG results, while the dashed curves correspond to Fokker--Planck simulations with $N_z=255$, $511$, and $1023$, respectively, using $N_\varphi=256$.}
    \label{fig:Nz analysis}
\end{figure*}

In contrast to Fokker--Planck analyses for ferromagnets, we find that the grid resolution, especially along the $x$ direction, is 
important for obtaining accurate numerical solutions in $\mathrm{Mn_3Sn}$. Because of the extremely strong perpendicular anisotropy, $\boldsymbol{m}_\mathrm{oct}$ is effectively confined to the $x$--$y$ plane. Nevertheless, the small but nonzero dynamics of its $z$ component remain essential in $\mathrm{Mn_3Sn}$. From a macroscopic viewpoint, when $\boldsymbol{m}_\mathrm{oct}$ tilts toward the $+z$ ($-z$) direction, a strong restoring torque is generated along the $-z$ ($+z$) direction. However, within the much narrower range $|z|<10^{-4}$, Fig.~\ref{fig:zdot} shows that positive (negative) $\dot{z}$ in the $+z$ ($-z$) half-space allows $\boldsymbol{m}_\mathrm{oct}$ to move naturally away from the $x$--$y$ plane. Although this out-of-plane excursion is extremely small, it is crucial for the ultrafast dynamics of $\mathrm{Mn_3Sn}$. As indicated by Eq.~\ref{eq: reduced}(b), $H_{\theta_\mathrm{oct}}$ also contributes to $\dot{\varphi}_\mathrm{oct}$, and its magnitude is approximately $5$--$6$ orders larger than that of $\alpha H_{\varphi_\mathrm{oct}}$. Consequently, even an out-of-plane deviation of $\boldsymbol{m}_\mathrm{oct}$ as small as $10^{-4}$ or less can significantly enhance $\dot{\varphi}_\mathrm{oct}$, as illustrated in Fig.~\ref{fig:zdot}. These results show that such tiny out-of-plane deviations cannot be neglected when modeling the ultrafast dynamics of $\mathrm{Mn_3Sn}$. This places a stringent requirement on the grid spacing in the numerical solution of the Fokker--Planck equation, which, according to our simulations, must be on the order of $10^{-6}$. Resolving the full interval $z\in[-1,1]$ at this resolution would require approximately $10^6$ grid points along the $z$ direction, which is computationally prohibitive with current resources.

To make the numerical simulation feasible with current computational resources, we restrict the simulation range of $z$ to the region where $P(z,t)$ is nonzero. We perform numerical Monte Carlo simulations of the reduced LLG equation and, as shown in Fig.~\ref{fig:zdot}(a), for $T=300~\mathrm{K}$ and an energy barrier of $\Delta E=2~k_\mathrm{B}T$, the distribution of $m_{\mathrm{oct},z}$ remains strictly confined within $|z|<10^{-3}$. Guided by these reduced LLG results, we therefore limit the computational domain to $z\in[-10^{-3},10^{-3}]$, which provides sufficient resolution over the physically relevant region while remaining computationally tractable. The results presented below further confirm that this truncated domain is adequate for accurate simulation and benchmarking. Under this choice, the usual mathematically guaranteed flux conservation at the poles ($z=\pm1$) no longer applies, because the computational boundaries are imposed artificially at $|z|=10^{-3}$. Instead, we enforce zero outward flux at these boundaries. This approximation does not compromise the physical accuracy of the simulation, since the reduced LLG results show that the probability of finding $\boldsymbol{m}_\mathrm{oct}$ outside this narrow interval is negligible.



In Fig.~\ref{fig:Nz analysis}, we further examine the effect of grid resolution by discretizing $z$ and $\varphi_\mathrm{oct}$ into $N_z$ and $N_\varphi=256$ grid points, respectively. In the deterministic limit, the Fokker--Planck results are benchmarked against reduced LLG simulations. Excellent agreement is obtained for all cases when $N_z=1023$. By contrast, smaller values of $N_z$ lead to visible discrepancies, particularly near $H_a=0.25~\mathrm{T}$. We therefore use $N_z=1023$ in all subsequent simulations. This analysis also highlights the impracticality of solving the Fokker--Planck equation over the full interval $-1\le z\le 1$ at comparable resolution: such a discretization would require approximately $2.6\times10^8$ grid elements, corresponding to about $2.8~\mathrm{GB}$ for a single double-precision matrix alone, without accounting for the substantially greater memory and computational costs associated with the matrix solver.

\begin{figure*}
    \centering
    \includegraphics[width=0.95\linewidth]{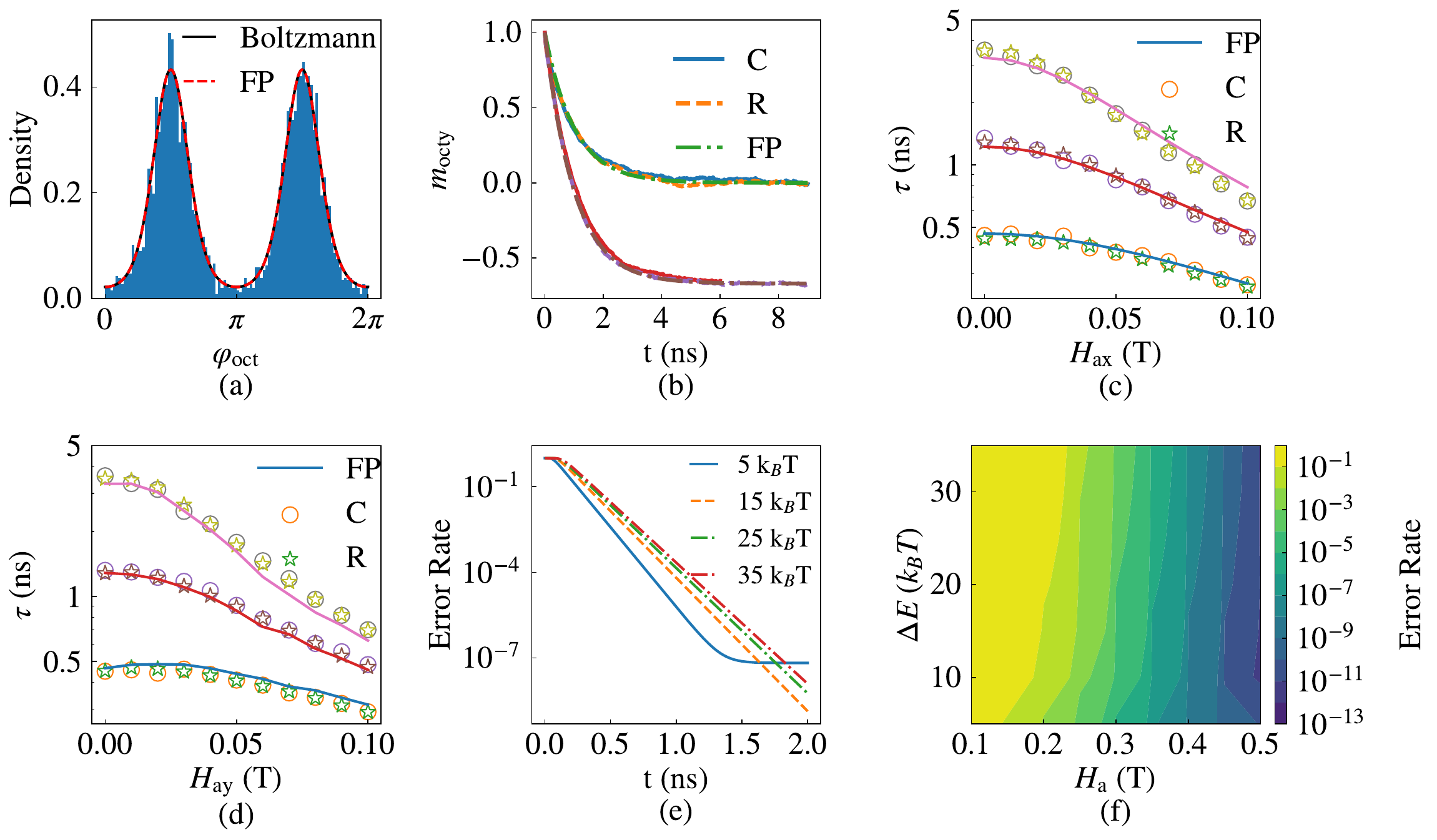}
    \caption{Validation and application of the reduced Fokker--Planck solver for thermally assisted switching in Mn$_3$Sn. (a) Equilibrium distribution of $\varphi_\mathrm{oct}$ for $\Delta E=3~k_\mathrm{B}T$ at $T=300~\mathrm{K}$ in the absence of an external field. The histogram denotes the distribution obtained from reduced LLG simulations, the red dashed curve shows the Fokker--Planck result, and the black solid curve represents the Boltzmann distribution. (b) Time evolution of $m_{\mathrm{oct},y}$ for $\Delta E=3~k_\mathrm{B}T$ and $H_a=0.5~\mathrm{T}$, comparing complete LLG (C), reduced LLG (R), and Fokker--Planck (FP) results for $\varphi_H=\pi$ and $\varphi_H=3\pi/2$. (c) Effective switching time $\tau$ as a function of $H_{ax}$ for the symmetric-field case $\varphi_H=\pi$, extracted from complete LLG, reduced LLG, and Fokker--Planck simulations. $\Delta E$ increases from 2~$k_\mathrm{B}T$ at the bottom to 4~$k_\mathrm{B}T$ at the top. (d) Effective switching time $\tau$ as a function of $H_{ay}$ for the symmetry-breaking case $\varphi_H=3\pi/2$, extracted from complete LLG, reduced LLG, and Fokker--Planck simulations. $\Delta E$ increases from 2~$k_\mathrm{B}T$ at the bottom to 4~$k_\mathrm{B}T$ at the top. (e) Error rate versus pulse width for field-assisted switching with $\varphi_H=3\pi/2$ at different energy barriers. (f) Error rate for a $1~\mathrm{ns}$ field pulse as a function of $\Delta E$ and $H_a$, showing that the reduced Fokker--Planck framework can efficiently access the ultra-low-error regime.}
    \label{fig:FP simulation}
\end{figure*}

Furthermore, we investigate the low-energy-barrier regime under weak external fields, $H_\mathrm{a} \leq 0.1~\mathrm{T}$, to assess the feasibility of external-field-assisted probabilistic switching. At equilibrium, the angular distribution follows the Boltzmann form,
\begin{equation}
P(\varphi_\mathrm{oct})=\lambda \exp[-\mathcal{H}(\varphi_\mathrm{oct})],
\end{equation}
where $\lambda$ is the normalization constant. Figure~\ref{fig:FP simulation}(a) shows the distribution of $\varphi_\mathrm{oct}$ for $\Delta E = 2~k_\mathrm{B}T$ at $T=300~\mathrm{K}$. In the absence of external stimuli, the distribution is symmetric about $\varphi_\mathrm{oct}=\pi$, as expected from the Boltzmann distribution. This behavior is confirmed by both the histogram obtained from reduced LLG simulations and the continuous distribution calculated from the Fokker--Planck solver.

We next apply an external field with $\varphi_H=\pi$ and $\varphi_H=3\pi/2$, respectively. Figure~\ref{fig:FP simulation}(b) shows the time evolution of $m_{\mathrm{oct},y}$ for $\Delta E = 3~k_\mathrm{B}T$ and $H_\mathrm{a}=0.5~\mathrm{T}$. The Fokker--Planck results agree closely with Monte Carlo simulations based on both the reduced and complete LLG equations, demonstrating the validity of the proposed Fokker--Planck solver. When $\varphi_H=\pi$, the symmetry of the Hamiltonian is preserved, so the two equilibrium states remain equivalent and $m_{\mathrm{oct},y,\infty}=0$. In this case, the effective switching time $\tau_0$ is extracted by fitting the relaxation curve to
\begin{equation}
m_{\mathrm{oct},y}(t)=\exp(-t/\tau_0).
\end{equation}
In contrast, when $\varphi_H=3\pi/2$, the external field breaks the symmetry between the two equilibrium states, and the state with $\varphi_\mathrm{oct}>\pi$ becomes increasingly favored as $H_\mathrm{a}$ increases. The corresponding relaxation is therefore fitted using
\begin{equation}
m_{\mathrm{oct},y}(t)=m_\infty + (1-m_\infty)\exp(-t/\tau_0).
\end{equation}

Figures~\ref{fig:FP simulation}(c) and \ref{fig:FP simulation}(d) summarize the extracted $\tau$ values from the Monte Carlo and Fokker--Planck simulations for these two field configurations. The good agreement further supports both the reduced LLG description and the corresponding Fokker--Planck formulation. 

More importantly, the Fokker--Planck method becomes particularly advantageous for analyzing storage devices or switching schemes operating in the ultra-low-error regime. To illustrate this capability, we further study field-assisted switching with $\varphi_\mathrm{H}=3\pi/2$ for $5k_\mathrm{B}T \leq \Delta E \leq 35k_\mathrm{B}T$ and $0.1~\mathrm{T} \leq H_\mathrm{a} \leq 0.5~\mathrm{T}$, as shown in Figs.~\ref{fig:FP simulation}(e) and \ref{fig:FP simulation}(f). The external field is applied to induce deterministic switching, such that $\boldsymbol{m}_\mathrm{oct}$ is expected to relax into the target energy well at $\varphi_\mathrm{oct} = 3\pi/2$. Since the switching probability is close to 100\% in this regime, we quantify the switching accuracy using the error rate, defined as the probability that $\boldsymbol{m}_\mathrm{oct}$ remains in the original state.

Before reaching steady state, the error rate decreases approximately exponentially with pulse width. A smaller $\Delta E$ leads to a faster decrease in the error rate because the switching dynamics are faster. As the system approaches equilibrium, however, the error rate saturates to a constant value. In this regime, a smaller $\Delta E$ produces a higher equilibrium error floor and reaches this equilibrium value more rapidly. Figure~\ref{fig:FP simulation}(e) shows an example for $H_\mathrm{a}=0.3~\mathrm{T}$. Within the simulated time window of $t=2~\mathrm{ns}$, only the $\Delta E = 5~k_\mathrm{B}T$ case reaches equilibrium. Figure~\ref{fig:FP simulation}(f) shows the error rate obtained for a $1~\mathrm{ns}$ pulse as a function of $\Delta E$ and $H_\mathrm{a}$, where the error probability can be as low as $10^{-13}$. Resolving such low error probabilities using direct Monte Carlo simulations would be computationally infeasible, as discussed below.

\begin{figure}
    \centering
    \includegraphics[width=0.95\linewidth]{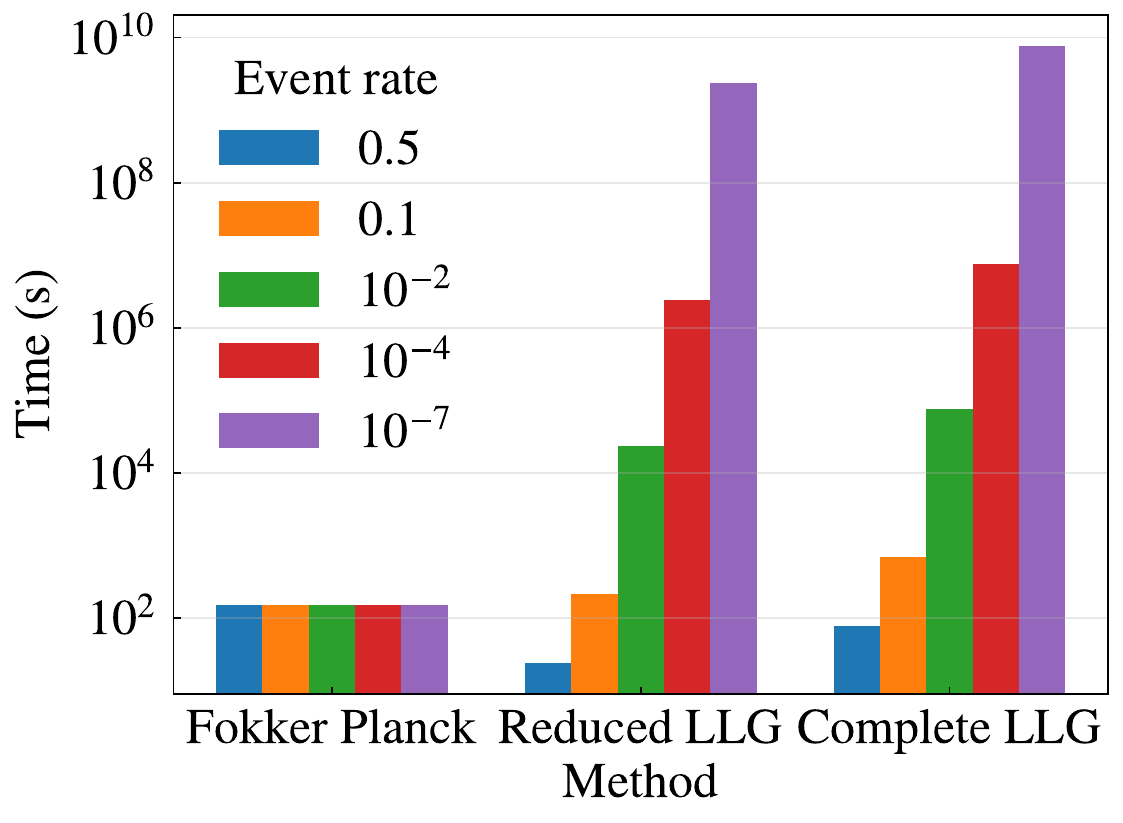}
    \caption{Runtime comparison for different event rates using the Fokker--Planck solver and Monte Carlo simulations.}
    \label{fig:run time}
\end{figure}

In our numerical Fokker--Planck solver, a 1-ns simulation with $\Delta t = 5~\mathrm{fs}$, $N_z = 1023$, and $N_\varphi = 256$ requires approximately $150$~s on an NVIDIA A30 GPU in double precision. For Monte Carlo simulations, although GPU parallelization provides substantial acceleration, the total runtime remains effectively linear in the ensemble size when the number of samples is sufficiently large, e.g., greater than $10^4$. Based on benchmarks performed on the same device, a 1-ns simulation with $\Delta t = 1~\mathrm{fs}$ requires $20~\mathrm{ms}$ per sample for the complete LLG equation and $6.26~\mathrm{ms}$ per sample for the reduced LLG equation. To obtain statistically reliable results, the required ensemble size increases rapidly as the event rate, such as the error rate, decreases. For example, achieving a $1\%$ relative error with a $95\%$ confidence interval requires approximately $3.5\times10^5$ samples for a $10\%$ event rate, whereas a $1\%$ event rate requires about $3.8\times10^6$ samples. Figure~\ref{fig:run time} compares the runtimes of the Fokker--Planck and Monte Carlo methods over event rates ranging from $0.5$ to $10^{-7}$. The Fokker--Planck approach begins to show a clear computational advantage when the event rate falls below $0.1$, while Monte Carlo simulation becomes nearly impractical for event rates below $10^{-4}$. These results strongly demonstrate that the Fokker--Planck method is an efficient and effective approach for the thermal analysis of $\mathrm{Mn_3Sn}$.


\vspace{-10pt}
\section{Conclusion}
\vspace{-10pt}
In this work, we developed a reduced LLG plus Fokker--Planck framework for thermally assisted octupolar dynamics in Mn$_3$Sn. The reduced LLG model was shown to reproduce the essential behavior of the complete three-sublattice dynamics, while the corresponding Fokker--Planck equation enabled efficient analysis of stochastic switching and relaxation. We found that accurate simulation requires extremely fine resolution in the out-of-plane coordinate because even very small deviations from the basal plane strongly influence the ultrafast dynamics. The reduced Fokker--Planck solver was validated against Monte Carlo simulations through equilibrium distributions, relaxation trajectories, and switching times, and it further enabled access to ultra-low error rates that are not practical to obtain by direct Monte Carlo sampling. These results establish the reduced stochastic framework as an effective approach for studying thermal switching and reliability in Mn$_3$Sn.

The ability to efficiently quantify thermally activated dynamics also makes this framework relevant for evaluating antiferromagnetic spintronic devices from an application-oriented perspective. In particular, access to switching-time distributions, retention behavior, and ultra-low error probabilities provides useful metrics for assessing the reliability of Mn$_3$Sn-based devices beyond deterministic switching characteristics alone. This capability is important for practical device design, where high-speed operation must be considered together with thermal stability and switching accuracy.

Building on this foundation, future work can further extend the framework to spin-current-driven dynamics, where thermal fluctuations may influence current-induced switching, oscillation, and device reliability. Such studies will be useful for clarifying the combined roles of spin-orbit torque, Joule heating, and thermal activation under practical operating conditions. In addition, further benchmarks against experimental measurements are expected, including comparisons with field-driven and current-driven switching data, relaxation behavior, and retention characteristics. These comparisons will help refine model parameters and further establish the predictive capability of the reduced Fokker--Planck framework for experimental Mn$_3$Sn devices.

\vspace{-10pt}
\section*{Acknowledgment}
\vspace{-10pt}
This work was supported by the National Science Foundation through the University of Illinois at Urbana-Champaign Materials Research Science and Engineering Center under Grant No.~DMR-1720633. The authors also acknowledge partial support from the National Science Foundation through the Center for Advanced Semiconductor Chips with Accelerated Performance Industry--University Cooperative Research Center under NSF Cooperative Agreement No.~EEC-2231625.

\vspace{-10pt}
\section*{Author Declarations}
\vspace{-10pt}
\subsection*{Conflict of Interest}
\vspace{-10pt}
The authors have no conflicts to disclose.

\noindent
\subsection*{Author Contributions}
\vspace{-10pt}
\textbf{Siyuan Qian}: Conceptualization (equal); Data curation (lead);
Formal analysis (lead); Investigation (lead); Methodology (lead);
Software (lead); Validation (lead); Writing -- original draft (lead);
Writing -- review \& editing (supporting).
\textbf{Shaloo Rakheja}: Conceptualization (lead); Funding acquisition
(lead); Project administration (lead); Resources (lead); Supervision
(lead); Writing -- original draft (supporting); Writing -- review \&
editing (lead).

\clearpage
\onecolumngrid
\setcounter{figure}{0}
\setcounter{table}{0}
\setcounter{equation}{0}
\renewcommand{\thefigure}{S\arabic{figure}}
\renewcommand{\thetable}{S\arabic{table}}
\renewcommand{\theequation}{S\arabic{equation}}

\section*{Supplementary Material}
\addcontentsline{toc}{section}{Supplementary Material}


\section{Derivation of the Reduced LLG equation}

\subsection{Lagrangian in terms of $\boldsymbol{m}_\mathrm{oct}$}

Based on the definition of $\boldsymbol{m}_\mathrm{oct}$, the following relationships hold:
\begin{subequations}
    \begin{equation}
        \varphi_\mathrm{oct} = -\frac{\varphi_1+\varphi_2+\varphi_3}{3},
    \end{equation}
    \begin{equation}
        \theta_\mathrm{oct} = \frac{\theta_1+\theta_2+\theta_3}{3},
    \end{equation}
\end{subequations}
where $\phi_j$ and $\theta_j$ ($j=1, 2, 3$) denote ....

Under the perturbative assumption that the sublattices exhibit only small deviations from the equilibrium state, $\varphi_i$ and $\theta_i$ can be written as:
\begin{subequations}
    \begin{equation}
        \varphi_i = -\varphi_\mathrm{oct}+\frac{2i}{3}\pi + \delta \varphi_i
    \end{equation}
    \begin{equation}
        \theta_i = \theta_\mathrm{oct}+\delta \theta_i
    \end{equation}
\end{subequations}
where $\delta \varphi_i$ and $\delta \theta_i$ are small deviations compared with $\varphi_i$ and $\theta_i$, by neglecting these small deviations, we approximate the relationships as:
\begin{subequations}
    \begin{equation}
        \dot{\varphi}_\mathrm{oct} \approx -\dot{\varphi}_1 \approx -\dot{\varphi}_2 \approx -\dot{\varphi}_3,
    \end{equation}
    \begin{equation}    
        \theta_\mathrm{oct} \approx \theta_1 \approx \theta_2 \approx \theta_3.
    \end{equation}
\end{subequations}

We now express $\mathcal{L}_\mathrm{B}$, $\mathcal{H}$, and $\mathcal{W}$ in terms of $\theta_\mathrm{oct}$ and $\varphi_\mathrm{oct}$, term by term.

\begin{equation}
    \begin{aligned}
        \mathcal{L}_\mathrm{B} = \sum_{i=1}^3 {\frac{M_\mathrm{s}}{\gamma} (1-\cos \theta_i) \dot{\varphi}_i}
        = - \frac{3M_\mathrm{s}}{\gamma}(1- \cos \theta_\mathrm{oct})\dot{\varphi}_\mathrm{oct}.
    \end{aligned}
\end{equation}

From existing perturbation theory results,~\cite{Qian2025,Konakanchi2025,shukla2024impact} the non-thermal terms of the Hamiltonian have already been expressed in terms of $\theta_\mathrm{oct}$ and $\varphi_\mathrm{oct}$. Therefore, we only need to treat the thermal field term. Since the thermal field follows a zero-mean Gaussian distribution, part of the sign convention is absorbed in the following expression:
\begin{equation}
    \begin{aligned}
        \mathcal{H}_{th} &= - \sum_{i=1}^3 M_\mathrm{s}\boldsymbol{m}_i \cdot \boldsymbol{H}_{th,i}
        = - M_\mathrm{s} \sum_{i=0}^3 \left(H_{x,i} \sin \theta_i \cos \varphi_i+H_{y,i} \sin\theta_i \sin\varphi_i+H_{z,i} \cos\theta_i\right) \\
        &= -M_\mathrm{s} \sum_{i=0}^3 \left(H_{x,i} \sin \theta_\mathrm{oct} \cos \varphi_\mathrm{oct}+H_{y,i} \sin\theta_\mathrm{oct} \sin\varphi_\mathrm{oct}+H_{z,i} \cos\theta_\mathrm{oct} \right) \\
        &= -M_\mathrm{s} \boldsymbol{m}_\mathrm{oct} \cdot (\boldsymbol{H}_{th,1}+\boldsymbol{H}_{th,2}+\boldsymbol{H}_{th,3})
        = -3M_\mathrm{s} \boldsymbol{m}_\mathrm{oct} \cdot \boldsymbol{H}_{th,\mathrm{oct}},
    \end{aligned}
\end{equation}
where $\boldsymbol{H}_\mathrm{th,oct} = \frac{1}{3}(\boldsymbol{H}_{th,1}+\boldsymbol{H}_{th,2}+\boldsymbol{H}_{th,3})$ has one-third of the correlation strength of $\boldsymbol{H}_\mathrm{th,i}$, since the three components follow independent Gaussian distributions.

The dissipation term, including damping and spin torque, is given by
\begin{equation}
    \begin{aligned}
        \mathcal{W} &= \sum_{i=1}^3 \frac{\alpha M_\mathrm{s}}{2\gamma}(\dot{\theta}_i^2 +\sin^2{\theta_i} \dot{\varphi}_i)
        + \frac{M_\mathrm{s}}{\gamma}\tau(\dot{\theta}_i \Gamma_\varphi(\varphi_i)+\sin \theta_i \dot{\varphi}_i \Gamma_\theta(\varphi_i)) \\
        &= \frac{3\alpha M_s }{2\gamma}\left(\dot{\theta}_\mathrm{oct}^2 + \sin^2\theta_\mathrm{oct}\,\dot{\varphi}_\mathrm{oct}^2\right)
        +\frac{3M_s }{\gamma}\,\tau\left(\dot{\theta}_\mathrm{oct}\,\Gamma_{\varphi}(-\varphi_\mathrm{oct})
        +\sin\theta_\mathrm{oct}\,\dot{\varphi}_\mathrm{oct}\,\Gamma_{\theta}(-\varphi_\mathrm{oct})\right).
    \end{aligned}
\end{equation}

\subsection{Derivation of reduced LLG equation from Lagrangian}

We now substitute the above expressions into the Lagrange equation. For the $\theta$ direction:
\begin{subequations}
    \begin{equation}
        \frac{\partial \mathcal{L}}{\partial \theta_\mathrm{oct}} = - \frac{3M_\mathrm{s}}{\gamma}\sin \theta_\mathrm{oct}\dot{\varphi}_\mathrm{oct} - \frac{\partial \mathcal{H}}{\partial \theta_\mathrm{oct}},
    \end{equation}
    \begin{equation}
        - \frac{d}{dt} \frac{\partial \mathcal{L}}{\partial \dot{\theta}_\mathrm{oct}} = 0,
    \end{equation}
    \begin{equation}
        \frac{\partial \mathcal{W}}{\partial \dot{\theta}_\mathrm{oct}} = \frac{3\alpha M_\mathrm{s}}{\gamma} \dot{\theta}_\mathrm{oct} + \frac{3 M_\mathrm{s}}{\gamma}\tau \Gamma_\varphi(-\varphi_\mathrm{oct}),
    \end{equation}
\end{subequations}
which leads to
\begin{equation}
    \alpha \dot{\theta}_\mathrm{oct} + \sin \theta_\mathrm{oct} \dot{\varphi}_\mathrm{oct}
    = -\frac{\gamma}{3 M_\mathrm{s}} \frac{\partial \mathcal{H}}{\partial \theta_\mathrm{oct}}
    - \tau \Gamma_\varphi(-\varphi_\mathrm{oct}).
\end{equation}

Similarly, for the $\varphi$ direction:
\begin{subequations}
    \begin{equation}
        \frac{\partial \mathcal{L}}{\partial \varphi_\mathrm{oct}} = -\frac{\partial \mathcal{H}}{\partial \varphi_\mathrm{oct}},
    \end{equation}
    \begin{equation}
        - \frac{d}{dt} \frac{\partial \mathcal{L}}{\partial \dot{\varphi}_\mathrm{oct}} = -\frac{3 M_\mathrm{s}}{\gamma} \sin \theta_\mathrm{oct} \dot{\theta}_\mathrm{oct},
    \end{equation}
    \begin{equation}
        \frac{\partial \mathcal{W}}{\partial \dot{\varphi}_\mathrm{oct}} = \frac{3\alpha M_\mathrm{s}}{\gamma} \sin^2 \theta_\mathrm{oct} \dot{\varphi}_\mathrm{oct} + \frac{3 M_\mathrm{s}}{\gamma} \tau \sin \theta_\mathrm{oct} \Gamma_\theta(-\varphi_\mathrm{oct}),
    \end{equation}
\end{subequations}
which yields
\begin{equation}
    \dot{\theta}_\mathrm{oct} + \alpha \sin^2 \theta_\mathrm{oct} \dot{\varphi}_\mathrm{oct}
    = -\frac{\gamma}{3M_\mathrm{s}\sin \theta_\mathrm{oct}} \frac{\partial \mathcal{H}}{\partial \varphi_\mathrm{oct}}
    - \tau \sin \theta_\mathrm{oct} \Gamma_\theta(-\varphi_\mathrm{oct}).
\end{equation}

Using the relations
\[
\boldsymbol{H}_\theta = -\frac{1}{3M_\mathrm{s}} \frac{\partial \mathcal{H}}{\partial \theta_\mathrm{oct}},
\qquad
\boldsymbol{H}_\varphi = -\frac{1}{3 M_\mathrm{s}\sin \theta_\mathrm{oct}} \frac{\partial \mathcal{H}}{\partial \varphi_\mathrm{oct}},
\]
we obtain the reduced LLG equation:
\begin{subequations}
    \begin{equation}
        \begin{aligned}
            (1+\alpha^2)\,\dot{\theta}_\mathrm{oct} = \gamma\left(\alpha H_{\theta_\mathrm{oct}}-H_{\varphi_\mathrm{oct}}\right) 
            + \tau\left(\Gamma_{\theta}(\varphi_\mathrm{oct}) - \alpha\,\Gamma_{\varphi}(-\varphi_\mathrm{oct})\right),
        \end{aligned}
    \end{equation}
    \begin{equation}
        \begin{aligned}
            (1+\alpha^2)\,\sin \theta_\mathrm{oct}\,\dot{\varphi}_\mathrm{oct} = \gamma\left(H_{\theta_\mathrm{oct}}+ \alpha H_{\varphi_\mathrm{oct}}\right) 
            - \tau\left(\alpha\,\Gamma_{\theta}(\varphi_\mathrm{oct}) + \Gamma_{\varphi}(-\varphi_\mathrm{oct})\right).
        \end{aligned}
    \end{equation}
    \label{supp:eq: reduced}
\end{subequations}

\section{Derivation of the Fokker--Planck Equation}

\subsection{Thermal-Field Correlations in Spherical Coordinates}

We start from the thermal field expressed in Cartesian coordinates,
\begin{equation}
\boldsymbol{H}^{\mathrm{th}}(t)
=
\bigl(
h_x(t),\,h_y(t),\,h_z(t)
\bigr),
\end{equation}
with zero mean and isotropic white-noise correlations:
\begin{equation}
\langle h_i(t)\rangle = 0,
\end{equation}
\begin{equation}
\langle h_i(t) h_j(t')\rangle
=
2D\,\delta_{ij}\,\delta(t-t'),
\qquad
i,j\in\{x,y,z\},
\label{supp:eq:cartesian_thermal_corr}
\end{equation}
where
\begin{equation}
D = \frac{2\alpha k_\mathrm{B} T}{3\gamma \mu_0 M_\mathrm{s}^2 \mathcal{V}}.
\end{equation}

We parameterize the octupolar order parameter direction using the spherical angles
$\theta_{\mathrm{oct}}$ and $\varphi_{\mathrm{oct}}$. The corresponding local orthonormal basis vectors are given by
\begin{equation}
\boldsymbol{e}_{r}
=
\left(
\sin\theta_{\mathrm{oct}}\cos\varphi_{\mathrm{oct}},
\sin\theta_{\mathrm{oct}}\sin\varphi_{\mathrm{oct}},
\cos\theta_{\mathrm{oct}}
\right),
\end{equation}
\begin{equation}
\boldsymbol{e}_{\theta_{\mathrm{oct}}}
=
\left(
\cos\theta_{\mathrm{oct}}\cos\varphi_{\mathrm{oct}},
\cos\theta_{\mathrm{oct}}\sin\varphi_{\mathrm{oct}},
-\sin\theta_{\mathrm{oct}}
\right),
\end{equation}
\begin{equation}
\boldsymbol{e}_{\varphi_{\mathrm{oct}}}
=
\left(
-\sin\varphi_{\mathrm{oct}},
\cos\varphi_{\mathrm{oct}},
0
\right).
\end{equation}

The thermal field components in spherical coordinates are obtained by projecting onto the local tangent basis:
\begin{equation}
h_{\theta_{\mathrm{oct}}}
=
\boldsymbol{H}^{\mathrm{th}}\cdot \boldsymbol{e}_{\theta_{\mathrm{oct}}},
\qquad
h_{\varphi_{\mathrm{oct}}}
=
\boldsymbol{H}^{\mathrm{th}}\cdot \boldsymbol{e}_{\varphi_{\mathrm{oct}}}.
\end{equation}

Explicitly, these components are
\begin{equation}
h_{\theta_{\mathrm{oct}}}
=
h_x \cos\theta_{\mathrm{oct}}\cos\varphi_{\mathrm{oct}}
+
h_y \cos\theta_{\mathrm{oct}}\sin\varphi_{\mathrm{oct}}
-
h_z \sin\theta_{\mathrm{oct}},
\end{equation}
\begin{equation}
h_{\varphi_{\mathrm{oct}}}
=
-h_x \sin\varphi_{\mathrm{oct}}
+
h_y \cos\varphi_{\mathrm{oct}}.
\end{equation}

Therefore, the thermal field components in spherical coordinates satisfy
\begin{equation}
\langle h_{\theta_{\mathrm{oct}}}(t)\rangle = 0,
\qquad
\langle h_{\varphi_{\mathrm{oct}}}(t)\rangle = 0,
\end{equation}
\begin{equation}
\left\langle
h_{\theta_{\mathrm{oct}}}(t)
h_{\theta_{\mathrm{oct}}}(t')
\right\rangle
=
\left\langle
h_{\varphi_{\mathrm{oct}}}(t)
h_{\varphi_{\mathrm{oct}}}(t')
\right\rangle
=
2D_h\,\delta(t-t'),
\end{equation}
\begin{equation}
\left\langle
h_{\theta_{\mathrm{oct}}}(t)
h_{\varphi_{\mathrm{oct}}}(t')
\right\rangle
=
0.
\end{equation}

Thus, isotropic white noise in Cartesian coordinates remains isotropic and uncorrelated in the local tangent-plane basis spanned by $\boldsymbol{e}_{\theta_{\mathrm{oct}}}$ and $\boldsymbol{e}_{\varphi_{\mathrm{oct}}}$.

\subsection{Full Expression of the Fokker--Planck Equation for $\boldsymbol{m}_\mathrm{oct}$}

We start from the stochastic octupolar dynamics,
\begin{subequations}
    \begin{equation}
        \begin{aligned}
            (1+\alpha^2)\,\dot{\theta}_\mathrm{oct} = \gamma\left(\alpha H_{\theta_\mathrm{oct}}-H_{\varphi_\mathrm{oct}}\right) 
            + \tau\left(\Gamma_{\theta}(\varphi_\mathrm{oct}) - \alpha\,\Gamma_{\varphi}(-\varphi_\mathrm{oct})\right),
        \end{aligned}
    \end{equation}
    \begin{equation}
        \begin{aligned}
            (1+\alpha^2)\,\sin \theta_\mathrm{oct}\,\dot{\varphi}_\mathrm{oct} = \gamma\left(H_{\theta_\mathrm{oct}}+ \alpha H_{\varphi_\mathrm{oct}}\right) 
            - \tau\left(\alpha\,\Gamma_{\theta}(\varphi_\mathrm{oct}) + \Gamma_{\varphi}(-\varphi_\mathrm{oct})\right).
        \end{aligned}
    \end{equation}
    \label{supp:eq: reduced stochastic}
\end{subequations}

The effective fields are decomposed into deterministic and stochastic parts,
\begin{equation}
H_{\theta_{\mathrm{oct}}}
=
H_{\theta_{\mathrm{oct}}}^{(0)} + h_{\theta_{\mathrm{oct}}}(t),
\qquad
H_{\varphi_{\mathrm{oct}}}
=
H_{\varphi_{\mathrm{oct}}}^{(0)} + h_{\varphi_{\mathrm{oct}}}(t).
\end{equation}

Accordingly, the Fokker--Planck equation for the probability density function $P(\theta_{\mathrm{oct}},\varphi_{\mathrm{oct}},t)$ is
\begin{subequations}
\begin{equation}
\frac{\partial P}{\partial t}
=
-\nabla\cdot \mathcal{J}
=
-\frac{1}{\sin\theta_{\mathrm{oct}}}\,
\partial_{\theta_{\mathrm{oct}}}
\left(
\sin\theta_{\mathrm{oct}}\,\mathcal{J}_{\theta_{\mathrm{oct}}}
\right)
-
\frac{1}{\sin\theta_{\mathrm{oct}}}\,
\partial_{\varphi_{\mathrm{oct}}}\mathcal{J}_{\varphi_{\mathrm{oct}}},
\label{supp:eq:fp_theta_phi}
\end{equation}
\begin{equation}
\mathcal{J}_{\theta_{\mathrm{oct}}}
=
\dot{\theta}_{\mathrm{oct}}\,P
-
D\frac{\partial P}{\partial \theta_{\mathrm{oct}}},
\qquad
\mathcal{J}_{\varphi_{\mathrm{oct}}}
=
\dot{\varphi}_{\mathrm{oct}}\,P
-
\frac{D}{\sin\theta_{\mathrm{oct}}}
\frac{\partial P}{\partial \varphi_{\mathrm{oct}}},
\label{supp:eq:fp_flux_theta_phi}
\end{equation}
\end{subequations}
where $\mathcal{J}$ is the probability flux. The diffusion constant is
\begin{equation}
D=\frac{\alpha \gamma k_B T}{3(1+\alpha^2)\,M_s \mathcal{V}},
\label{supp:eq:diffusion_const_oct}
\end{equation}
which is reduced by a factor of $1/3$ relative to the ferromagnetic case.

\subsection{Transformations prepared for numerical solve}

To avoid singularities in the numerical solution near the poles and to enable a uniform grid in $\cos \theta_\mathrm{oct}$, we introduce the following transformation:
\begin{equation}
z=\cos\theta_{\mathrm{oct}},
\qquad -1\le z\le 1.
\label{supp:eq:z_def}
\end{equation}
Under this transformation,
\begin{equation}
\sin\theta_{\mathrm{oct}}=\sqrt{1-z^2},
\qquad
\frac{dz}{d\theta_{\mathrm{oct}}}
=
-\sin\theta_{\mathrm{oct}},
\end{equation}
so that
\begin{equation}
\frac{\partial}{\partial \theta_{\mathrm{oct}}}
=
-\sqrt{1-z^2}\,\frac{\partial}{\partial z}.
\end{equation}

The Fokker--Planck equation can then be written as
\begin{equation}
    \frac{\partial P}{\partial t}
    = \frac{\partial}{\partial z}
      \left[\sqrt{1-z^{2}}\,A_{\theta}\,P
            + D(1-z^{2})\frac{\partial P}{\partial z}\right]
    - \frac{1}{\sqrt{1-z^{2}}}\frac{\partial}{\partial\varphi}
      \left[A_{\varphi}P
            - \frac{D}{\sqrt{1-z^{2}}}\frac{\partial P}{\partial\varphi}\right].
\end{equation}

A purely explicit numerical scheme requires an extremely small time step, $\Delta t \leq \frac{\Delta z^2}{D}$, to maintain stability, whereas a fully implicit scheme requires greater computational resources. Therefore, we solve the drift term explicitly and the diffusion term implicitly. To this end, we first rewrite the equation as
\begin{equation}
\frac{\partial P}{\partial t}
=
\mathcal{A}(P)+\mathcal{D}(P),
\end{equation}
where
\begin{equation}
\mathcal{A}(P)
=
-\partial_z(\dot{z}P)
-\frac{1}{\sqrt{1-z^2}}\,
\partial_{\varphi_{\mathrm{oct}}}
(\dot{\varphi}_{\mathrm{oct}}P)
\end{equation}
is the deterministic drift operator, and
\begin{equation}
\mathcal{D}(P)
=
\partial_z\!\left(D(1-z^2)\partial_zP\right)
+
\partial_{\varphi_{\mathrm{oct}}}\!\left(
\frac{D}{1-z^2}\partial_{\varphi_{\mathrm{oct}}}P
\right)
\end{equation}
is the diffusion operator.

Using a first-order operator-splitting scheme, we evaluate the drift term explicitly:
\begin{equation}
\left(I-\Delta t\,\mathcal{D}\right)P^{n+1}
=
P^n+\Delta t\,\mathcal{A}(P^n)
\equiv P^{*}.
\end{equation}
We then apply a Fourier transform along $\varphi_{\mathrm{oct}}$, under which
\begin{equation}
\frac{\partial^2 P}{\partial \varphi_{\mathrm{oct}}^2}
\rightarrow
-m^2 P_m,
\end{equation}
where $P_m$ denotes the $m$th Fourier mode. With the Fourier transform applied, the implicit diffusion step is reduced to a tridiagonal system in $z$ for each mode $m$:
\begin{equation}
\begin{aligned}
P^{n+1}_{i,j,m}
&-\frac{D\Delta t}{\Delta z^2}
\Big[
(1-z_{i+1/2}^2)\left(P^{n+1}_{i+1,j,m}-P^{n+1}_{i,j,m}\right) \\
&\qquad
-(1-z_{i-1/2}^2)\left(P^{n+1}_{i,j,m}-P^{n+1}_{i-1,j,m}\right)
\Big]
+\frac{D\Delta t\,m^2}{1-z_i^2}P^{n+1}_{i,j,m}
=
P^*_{i,j,m}.
\end{aligned}
\end{equation}

After solving for $P^{n+1}_{i,j,m}$ for each mode, we perform the inverse Fourier transform to obtain $P^{n+1}_{i,j}$.

\end{document}